\documentstyle[multicol,prl,aps,epsf,psfig]{revtex}
\begin{document}

\title  
{
Intermittent Granular Flow and Clogging with Internal Avalanches
}
\author 
{
S. S. Manna$^{1,2}$ and H. J. Herrmann$^{1,3}$
}
\address
{
$^1$P. M. M. H., \'Ecole Sup\'erieure de Physique et Chimie Industrielles, 
10, rue Vauquelin, 75231 Paris Cedex 05 France \\ 
$^2$Satyendra Nath Bose National Centre for Basic Sciences,
Block-JD, Sector-III, Salt Lake, Calcutta 700 091, India \\
$^3$ICA1, University of Stuttgart, Pfaffenwaldring 27, 70569 Stuttgart, Germany \\
}
\maketitle
\begin{abstract}

The dynamics of intermittent granular flow through an orifice in a granular bin and the
associated clogging due to formation of arches blocking the outlet, is studied
numerically in two-dimensions. Our numerical results indicate that for small hole sizes,
the system self-organizes to a steady state and the distribution of the grain
displacements decays as power laws. On the other hand, for large holes, the
outflow distributions are also observed to follow power law distributions.

PACS number(s): 05.70.Jk, 64.60 Lx, 74.80.Bj, 46.10.+z

\end{abstract}

\begin{multicols}{2}

How much granular mass is expected to flow out through a hole at the bottom of a
granular bin? If the hole is sufficiently small, the outflow of grains stops soon due to the 
formation of a stable arch of grains clogging the hole. If the arch is broken, 
the grains in the arch become unstable and a cascade of grain displacements
propagates within the system called the `internal avalanche'. As a result,
a fresh flow starts through the hole, 
which also eventually stops due to the formation of another arch. It is known that 
the amount of out flowing granular mass between successive clogging events varies 
over a wide range, however, on average it depends sensitively on the hole 
diameter as compared to the grain size. 

In this paper, we study this phenomenon of clogged outflow distribution
from a granular bin in two-dimensions using computer simulations. We 
observe the power law dependence of the distribution. 

This phenomenon is important in the context of recent experimental
and theoretical studies of the avalanche size distributions from the
surfaces of a sandpile on an open base
\cite {btw,book,soc,chicago,ibm,oslo}. Research in this direction is 
motivated by the expectation that a sandpile is a simple example of a
Self-Organized Critical (SOC) system \cite {btw,book,soc}. According to the ideas of SOC,
long range spatio-temporal correlations may spontaneously emerge in 
slowly driven dynamical systems without fine tuning of any control parameter
\cite {btw}.

  Laboratory experiments on sandpiles, however, have not always found
evidence of criticality in sandpiles. The fluctuating outflow of granular 
mass from a slowly rotating horizontal semicircular drum did not show a power
law decay of its power spectrum as expected from the SOC theory \cite {chicago}.
In a second experiment, the avalanche size was directly measured by the outflow 
amount during the dropping of sand on a sandpile situated on a horizontal base.
It was observed that the distribution of the avalanche size, obeys a scaling 
behaviour only for small piles but not for the large ones \cite{ibm}.
However, it was claimed that a stretched exponential distribution for the avalanches
fits the whole range of system sizes \cite {feder}.
The reasons for not observing scaling are the existence of two angles of repose
as observed in \cite {chicago} and also the effect of inertia of 
the sand mass while moving down during avalanches. This effect
was suppressed in an experiment with anisotropic rice grains and scaling
was successfully observed \cite {oslo}.

%########################################################################
\begin{figure}
\begin{center}
\centerline{\epsffile{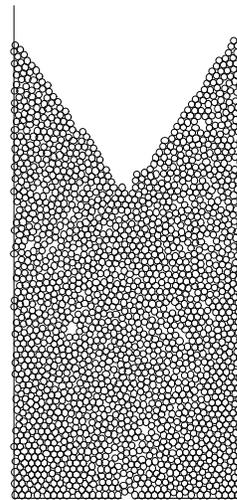}}
\narrowtext{\caption{
A granular system of 2000 grains of radii $1/\sqrt 2$ unit in a bin of width
50 units. The bin has a hole of width 3 units at the centre. The figure 
shows the steady state of the system after 3000 avalanches.
}}
\end{center}
\end{figure}
%########################################################################
\vskip -1.0 cm

  In contrast, we believe, internal avalanches are better candidates for
observing SOC. Due to the high compactness of grains in the granular material
in a bin, a grain never gets sufficient time to accelerate much and the
effect of inertia should be small. Internal avalanches were first studied 
in a block model of granular system by removing grains situated at the 
bottom of the bin one after another \cite {ball}. While this model and also 
other cellular automata \cite {disc} and `Tetris-like' models 
\cite {tetris} did exhibit scaling behaviour for the internal avalanches, the disc models 
\cite {disc,bang,manna} in continuous space, however, did not show 
sufficient evidence of SOC.

We study a two-dimensional granular system with grains kept in a vertical
rectangular bin. Initially the system has no hole at the bottom and we 
fill the bin up to a certain height. A hole of half-width $R$
is then made at the centre of the bottom. Some grains drop out through the hole. This
flow is stopped when an `arch' is formed, clogging the flow (Fig. 1). In two-dimensions,
an arch is a sequence of grains, mutually depending on one another,
such that their weights are balanced by mutual reaction forces.
These arches are formed due to the competition among the grains
to occupy the same vacant region of space.
In the steady state, these arches are observed everywhere.
The arch, covering the hole, is broken by deleting randomly one of the
two grains at the bottom of the arch. An avalanche of grain movements 
starts within the system. Those grains which reach the bottom
line with their centres within the hole are considered to flow out of the system
and are deleted. A deleted grain is immediately replaced again randomly on the top
surface. We measure the number of grains $\rho$ that flow out of the system 
during an avalanche between two successive clogging events and
study its distributions for various values of the hole sizes $R$.

  The granular system is represented by $N$ hard mono disperse discs of radii $r$. No
two grains are allowed to overlap but they can touch each other and one can
roll over the other if possible. A rectangular area of
size $L_x \times L_y$ on the $x-y$ plane represents our two dimensional bin.
Periodic boundary conditions are used along the $x$ direction and
gravity acts along the $-y$ direction. 

   A `pseudo-dynamics' \cite {disc} method is used to study the avalanches.
In this method we do not solve the classical equations of motion for the          
grain system as should have been done in a Molecular Dynamics method. 
Here, the direction of gravity and the local geometrical
constraints due to the presence of other grains govern the movement
of a grain. To justify the use of the pseudo-dynamics
we argue that due to the high compactness of the system
a grain never gets sufficient time to accelerate to large velocities. 
During the avalanches, the grains move in parallel. This is mimicked in the
simulation by discretizing the time. In one time unit, all grains are given
one chance each to move, by selecting them one after another in any
random sequence. A grain can move only in two ways: Either
it can $fall$ a certain distance or $roll$ a certain angle over another
grain. We used only one parameter $\delta$ to characterize these two cases.
Therefore $\delta$ is the maximum possible free fall a grain can have in
an unit time. While rolling, the centre of a grain moves along the arc of a circle of radius $2r$
on another grain in contact, the maximum length of the arc is also limited
to $\delta$. Therefore, the maximum angle a grain can roll freely is:
$\theta=\delta/2r$ over another grain in contact.
We believe this dynamics models an assembly of very 
light grains in $0+$ gravity. 

  The presence of other grains in the neighbourhood of a particular grain imposes
severe restriction to its possible movements. This is taken care off in the simulation 
using a search algorithm which digitizes the bin into a square grid
and associates the serial numbers of grains to the primitive cells of the grid.
A cell contains at most one grain when $r \ge 1/\sqrt 2$ is chosen.
Then it needs to search only 24 nearest neighbouring cells for
possible contact grains. A grain $n$ in the stable position is supported by
two other grains with serial numbers $n_L$ and $n_R$ on the left
$(L)$ and right $(R)$. The position of the grain $n$ is then updated according to the
following prescription:
\begin {itemize}
\item [$\bullet$] If $n_L=n_R=0$, it falls,
\item [$\bullet$] if $n_L \ne 0$ but $n_R = 0$, it rolls over $n_L$,
\item [$\bullet$] if $n_L = 0$ but $n_R \ne 0$, it rolls over $n_R$,
\item [$\bullet$] if $n_L \ne 0$ and $n_R \ne 0$ it is stable.
\end {itemize}

   When the grain $n$ with the centre at $(x_n,y_n)$ is allowed to fall,
it may come in contact with a grain $m$ at $(x_m,y_m)$ within the distance
$\delta$. Therefore, during the fall, the coordinates are updated as:
\[
x_n'=x_n \quad {\rm and} \quad
y_n'=y_m+\sqrt{4r^2-(x_n-x_m)^2} \quad 
\]
\[
{\rm if} \quad y_n-y'_n < \delta \quad
{\rm otherwise,} \quad y'_n=y_n-\delta.
\]
Similarly the grain $n$ may roll an angle $\theta'$ over a grain $t$ and 
come in contact with another grain $m$ at $(x_m,y_m)$ . The coordinates
of the grain $n$ in the new position where it touches both $t$ and $m$ are:
\[
x_n' = \frac{1}{2}(x_m+x_t) \pm
(y_m-y_t)\sqrt{\frac{4r^2}{d_{mt}^2}-\frac{1}{4}}
\]
\[
y_n' = \frac{1}{2}(y_m+y_t) \mp
(x_m-x_t)\sqrt{\frac{4r^2}{d_{mt}^2}-\frac{1}{4}}.
\]
Here $d_{mt}$ is the distance between the centres of the grains $m$ and $t$
and the $\pm$ signs are for the left and right rolls.
This position is accepted only if $\theta' \le \theta$ otherwise
the grain $n$ rolls down a maximum angle $\theta$ over the grain $t$.

  The initial configuration of grains within the bin is done by releasing grains
sequentially one after another along vertical trajectories with randomly
chosen horizontal positions. They fall till they touch the growing heap 
and are then allowed to roll down on the surface to their local stable 
positions. This is called the
`ballistic deposition with re-structuring method' (BDRM) \cite {meakin}. The initial pattern 
does not have arches since a grain while rolling down along the surface does 
not need to compete with any other grain. However, the arches are observed
when the system is disturbed by opening the outlet and allowing the grains
to flow out. 

  After a large number of out-flows, the system reaches a steady state.
One can characterize this state in many ways.
We characterize the steady state by the distribution of the angles of the
contact vectors. When two grains are in contact, we draw two vectors
from their centres to the contact points. The angle $\phi$ is the angle
between the contact vector and the $+x$ direction. We measured the 
probability distributions $P(\phi)$ in the range $0^o-360^o$ for the
initial state as well as in the steady state. To arrive at the steady 
state we discard $2N$ outflows, though the transition takes place
even earlier than that.

%########################################################################
\begin{figure}
\begin{center}
\epsfxsize=3.3in
\epsffile{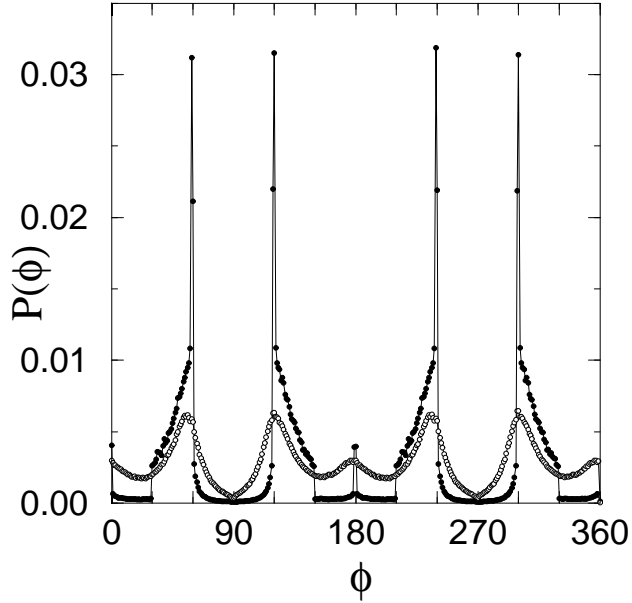}
\narrowtext{\caption{
Distribution $P(\phi)$ of the angles $\phi$ of the contact vectors.
The plot for the initial distribution is shown by black
dots. It has four high peaks at $60^o, 120^o, 240^o, 300^o$. The curve with
opaque circles corresponds to the steady state. The randomness introduced
into the system is reflected by the drastic reduction of the height of the
peaks as well as a broadening.
}}
\end{center}
\end{figure}
%########################################################################
\vskip -1.0 cm

  The contact angle distributions are shown in Fig. 2. The plot of 
$P(\phi)$ vs. $\phi$ for the initial grain distribution shows four sharp peaks at 
$60^o, 120^o, 240^o, 300^o$. We explain them in the following way.
If the grains were placed at the bottom in a complete orderly way by
mutually touching one another without any gap, the structure generated
by the BDRM method would be the hexagonal close packing (HCP) structure where
every grain has six other grains in contact at the interval of $60^o$s.
However, in our case, though we dropped grains randomly at the bottom level,
the deterministic piling process retains some effect of the HCP structure.
On the other hand, it is also known that the probability of a grain
having the number of contact neighbours different from four, decreases with the height as a
power law \cite {meakin}, which is certainly consistent with the four peaks we obtained.
There are also two small peaks at $0^o$ and at $180^o$ due
to the horizontal contacts of grains near the base level. This     
distribution, however, changes drastically by reducing the heights of the peaks and 
by broadening their widths in the steady state.
The randomness introduced into the system by randomly selecting 
grains at the bottom increases the possibilities of other contact angles
also, but is not fully capable to totally randomize it to a flat distribution.
We check that after a large number of avalanches, this steady state distribution
remains unchanged.

%########################################################################
\begin{figure}
\begin{center}
%\vskip -5.5 cm
\epsfxsize=3.3in
\epsffile{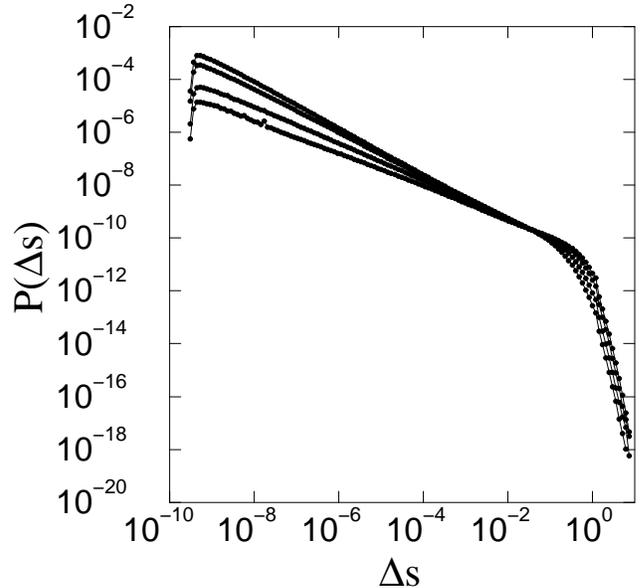}
%\vskip -2.0cm
\narrowtext{\caption{
Distribution $P(\Delta s)$ of the displacement $\Delta s$ of the grains
in a granular system in the limit of $R \rightarrow 0$.
Curves for different sizes with grain numbers $N$ = 300, 625, 2500 and
10000 are plotted from bottom to top on the left.
}}
\end{center}
\end{figure}
%########################################################################

  Next we study the statistics of the displacement distributions in the
granular heap in the limit of $R\rightarrow r$. In this limit, no grain
can drop out through the hole. The arch is broken by deleting only one
grain. The avalanche is allowed to propagate within the system. The
displacement $\Delta s$ is the absolute magnitude of the displacement
of a grain before and after an avalanche. We observe that there is a huge 
variation the displacements $\Delta s$ of the grains. Most of the grains 
displace very little, whereas others have much bigger displacements, but 
their numbers are small. The displacement distribution $P(\Delta s)$ is 
measured in the steady state over a large number of avalanches.
The lower cut-off of the distribution turned out to be strongly 
dependent on the tolerance factor $\epsilon$ used for the simulation.
If the centres of two grains are separated by a distance within
$r-\epsilon$ and $r+\epsilon$ then they are considered to be in contact.
The upper cut-off of the distribution is of the order of unity since
when a grain is deleted at the bottom, its neighbouring grains drops
a distance of the order of the grain diameter.

  Systems of four different sizes have been simulated with $N$ = 300, 625,
2500 and 10000 in bins with base sizes $L_x$ = 20, 40, 80 and 160
units respectively. First $5N$ avalanches are thrown away
to allow the system to reach the steady state. A list is maintained
for the coordinates of all grains. This list is compared before and
after the avalanche to measure $\Delta s$. We observe a nice straight 
line curve over many decades of the
$P(\Delta s)$ vs. $\Delta s$ plot on a double logarithmic scale, with
little curvature at the right edge (Fig. 3). However, the slope of this curve 
is found to increase with the system size. Different extrapolations are
tried and the best is obtained with $L^{-1/2}$ which leads to 0.97 $\pm$ 0.05 
in  the $N \rightarrow \infty$ limit for the slope.
We get similar behaviours for the distributions of the $x$ and $y$
components of the displacement vectors.

%########################################################################
\begin{figure}
\begin{center}
\epsfxsize=3.3in
\epsffile{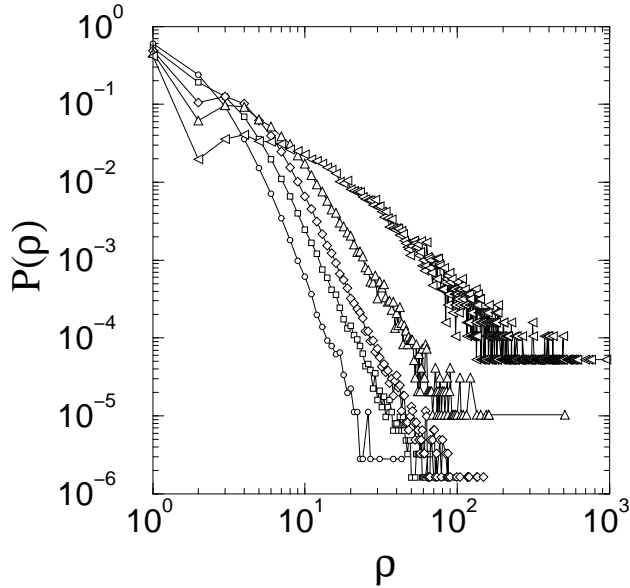}
\narrowtext{\caption{
The distribution $P(\rho)$ of the number of grains $\rho$ that flows out of the system in an
internal avalanche between two successive clogged situations
plotted on a double logarithmic scale. A bin of size $Lx=50$, containing
4000 grains of radii $r=1/\sqrt 2$ was simulated with the hole sizes $R$ = 0.75, 1, 1.25, 1.5, and 2.0
(curves from left to right).
}}
\end{center}
\end{figure}

%########################################################################
\vskip -0.5 cm

  Finally we study the outflow distribution and clogging.
We use a bin of dimensions $L_x=50$ and $L_y=200$ units containing
4000 grains of radii $r=1/\sqrt 2$ with a 
hole of half-width $R$ at the centre of the bottom of the bin.
In a stable   state an arch covers the hole and forbids other
grains to flow out. The two lower most grains of the arch on the two sides of 
the hole are chosen   and one of them is selected randomly.
By deleting this grain, the arch is made unstable, which creates an internal 
avalanche of grain movements within the bin.
Grains which come down and touch the bottom
line of the bin with their centres within the hole are
removed. While removing such a grain, the whole avalanche
dynamics is frozen and the grains is replaced back on the top
surface again by the BDRM method. 

  Two conical shapes are formed. There are some grains which are
never disturbed by any avalanche and are located on both sides
of the hole. These undisturbed grains form a cone.
The second cone is formed on the upper surface, though the grains
were dropped on this surface along randomly chosen horizontal
coordinates with uniform probability. For a fixed size of the base,
the angle of the cone is found to increase with the
average height of the granular column.

  The number of grains $\rho$ that go out of the system in an internal avalanche is counted
between successive clogging events. Holes of five different
sizes, i.e., $R = 0.75, 1, 1.25, 1.5, 2$ are used. We could generate 
around $5 \times 10^5$ such outflows for each hole size 
and collected the distribution data for $P(\rho)$.
We show the plot of these curves in Fig. 4 on a double logarithmic scale.
The grain which is deleted to break the arch is also counted in $\rho$.
It is found increasingly improbable that the outflow will consist of
only another grain. Consequently $P(1)$ is found to decrease monotonically
on increasing $R$. However, for large values of $\rho$, the curve falls
down.

  All curves of $P(\rho)$ vs. $\rho$ plots show straight lines for large
$\rho$ values, signifying the power law distributions for the avalanche sizes.
The slopes of these curves $\sigma (R)$ are found to depend strongly on the  
size of the hole, as well as the grain size: $P(\rho) \sim \rho^{-\sigma(R,r)}$.
We observe that $\sigma(R,r)$ is actually a linear function of $(R/r)$
as $\sigma(R,r)=k(\frac{R}{r})$ with k=1.6.
The average outflow $<\rho(R/r)>$ diverges exponentially as:
$exp(\alpha(\frac{R}{r}-1))$ with $\alpha = 0.7$.

  To summarize, we have studied a numerical model of intermittent granular flow
and clogging with associated internal avalanches from a two-dimensional 
bin using a pseudo dynamics method. The flow of grains through a hole at the 
bottom of the bin are repeatedly clogged due to the formation of stable arches of 
grains, which we break to start fresh flows. Our numerical simulation results
indicate that the system self-organizes to a state different from the initial
state. In the limit of $R \le r$, the distribution of granular distributions
follow power laws. On the other hand when $R > r$, the granular flow distribution
is also observed to follow power law.

  Funding by the Indo-French Centre for the Promotion of
Advanced Research (IFCPAR) through a project is thankfully acknowledged.

\leftline {Electronic Address: manna@boson.bose.res.in}
\leftline {Electronic Address: hans@ica1.uni-stuttgart.de}

%########################################################################

\end{multicols}

\end {document}